\documentclass{amsart}
\usepackage{amssymb, amsmath, epsfig}
\usepackage{amscd}

\newtheorem{thm}{Theorem} 
 
\newtheorem{lem}[thm]{Lemma}

\newtheorem{definition}{Definition}

\newcommand{\R}{\mathbb{ R}} 
\newcommand{\h}{\mathfrak{ h}} 
\newcommand{\g}{\mathfrak{ g}} 
\newcommand{\D}{\mathfrak{ d}} 
\newcommand{\N}{\mathcal{ N}} 
\newcommand{\T}{\mathcal{ T}}

\newcommand{\I}{\mathcal I} 
\newcommand{\q}{\mathbf q} 
 
\DeclareMathOperator{\pr}{pr} 
\DeclareMathOperator{\Span}{span} 
\DeclareMathOperator{\diag}{diag}

\title[Quasi-Chaplygin Systems]{Quasi-Chaplygin Systems and Nonholonimic Rigid 
Body Dynamics\footnote{Journal Ref:  Letters in Mathematical Physics (2006)}}

\author{Yuri N. Fedorov \and Bo\v zidar Jovanovi\' c}

\address{Department of Mathematics and Mechanics,Moscow Lomonosov University, Moscow, 119 899, Russia, and
Departament de Matem\`atica I, Universitat Politecnica de Catalunya, Barcelona, E-08028 Spain}
\email{fedorov@mech.math.msu.su,  Yuri.Fedorov@upc.es}

\address{Mathematical Institute SANU, Kneza Mihaila 35, 11000, Belgrade, Serbia and Montenegro}
\email {bozaj@mi.sanu.ac.yu}

\begin{document}  
\maketitle 
 
\begin{abstract} We show that the Suslov nonholonomic rigid body problem 
studied in \cite{FeKo,Jo4,Bl_Z} can be 
regarded almost everywhere as a generalized Chaplygin system. 
Furthermore, this provides a new example of a multidimensional nonholonomic system which can 
be reduced to a Hamiltonian form by means of Chaplygin reducing multiplier. 
Since we deal with Chaplygin systems in the local sense, the invariant manifolds of the 
integrable examples are not necessary tori. 
\end{abstract} 
 
\section{Introduction} 
 
We start the paper with the definition of nonholonomic Chaplygin systems, 
their reductions and Hamiltonization. 
 
\subsection*{Chaplygin Systems} 

Suppose we are given a natural nonholonomic system on the $n$-dimensional 
Riemannian manifold $(N,\varkappa)$ with local coordinates $x_i$, 
Lagrangian $l(x,\dot x)=\frac12 \sum\varkappa_{ij}\dot x_i\dot x_j-v(x)$ and 
$k$-dimensional distribution $D\subset TN$ 
describing kinematic constraints: a curve $x(t)$ is said to satisfy the 
constraints if $\dot x(t)\in D_{x(t)}$ for all $t$. 
The trajectory of the system $x(t)$ that satisfies the constraints 
is a solution to the Lagrange--d'Alembert equation 
\begin{equation} 
\sum_{i=1}^n \left(\frac{\partial l}{\partial x_i} 
- \frac{d}{dt}\frac{\partial l}{\partial \dot x_i}\right)\eta_i=0, 
\quad \mathrm{for\;all} \quad \eta\in D_x.  \label{Lagrange} 
\end{equation} 
 
Assume that $N$ has a principal bundle structure  $\pi: N\to 
Q=N/{\mathfrak G}$  with respect to the {\it left\/} action of a 
$(n-k)$-dimensional Lie group ${\mathfrak G}$ and $D$ is the 
collection of horizontal spaces of a principal connection. Given 
a vector $X_x\in T_x N$, we have the decomposition 
$X_x=X_x^h+X_x^v$, where $X_x^h\in D_x$, $X_x^v\in V_x$. Here 
$V_x$ is tangent space to the fiber $\mathfrak G\cdot x$ (the 
vertical space at $x$). 
 
Further, suppose that the Lagrangian $l$ is also $\mathfrak G$-invariant, i.e., 
$v$ is a $\mathfrak G$-invariant function 
and $\mathfrak G$ acts by isometries on Riemannian manifold $(N,\varkappa)$. 
Then the {\it constrained Lagrangian} $l_c(x,\dot x)=l(x,\dot x^h)$ 
induces a well defined reduced Lagrangian $L: TQ\to\R$ via identification 
$TQ\approx D/{\mathfrak G}$. 
The reduced Lagrangian $L$ is of the natural mechanical type as well, 
the corresponding kinetic energy (metric) and the potential energy 
will be denoted by $\varkappa_D$ and $V$ respectively. 
 
As a result, equation (\ref{Lagrange}) is $\mathfrak G$-invariant and 
defines a reduced  Lagrange--d'Alembert system on the 
tangent bundle $TQ$ (for the details see \cite{Koi, BKMM, CCLM}). 
After the Legendre transformation it can be rewritten as the following 
first-order dynamical system on $T^*Q$ 
\begin{equation} 
\dot q_i=\frac{\partial H}{\partial P_i},\quad \dot 
P_i=-\frac{\partial H}{\partial q_i}+\Pi_i(q,P) \qquad 
i=1,\dots,k, \label{2.1} 
\end{equation} 
where $q=(q_1,\dots,q_k)$ are local coordinates on the base space 
$Q$ and $P_i=\partial L/\partial \dot q_i$, $i=1,\dots,k$ are 
conjugate momenta. The Hamiltonian $H=\frac12 \sum 
\varkappa_D^{ij} P_iP_j+V(q)$ is the Legendre transformation of 
the reduced Lagrangian $L=\frac12\sum {\varkappa_D}_{ij} \dot 
q_i\dot q_j -V(q)$. The functions $\Pi_i$ are quadratic in momenta 
and depend on the curvature of the principal connection and the 
metric $\varkappa$. 
 
The system $(N,l,D,\mathfrak G)$ is called {\it a (generalized) 
Chaplygin system} (see Koiller \cite {Koi}), as a generalization 
of classical Chaplygin systems with Abelian symmetries \cite{Ch}. 
 
\subsection*{Chaplygin Reducing Multiplier} 

Let ${\mathcal N}(q)$ be a differentiable nonvanishing function on 
$Q$. Then, under the time substitution $d\tau={\mathcal N} (q)dt$ the 
following commutative diagram holds 
\begin{equation*} 
\begin{CD} 
 TQ\{q,\dot q\} @ > q^{\prime}=\dot q/{\mathcal N}(q) >> TQ\{q,q^{\prime}\} \\ 
 @ V P=\varkappa_D \dot q VV   @ VV p={\mathcal N}^2 \varkappa_D q^{\prime} V \\ 
 T^*Q\{q,P\}  @ >  p=\mathcal{\ N} P >>  T^*Q\{q,p\} . 
\end{CD} 
\end{equation*} 
Here $q^{\prime}={dq}/{d\tau}$. 
In the coordinates $\{q,q^{\prime}\} $ and $\{q,p\}$, $L$ and $H$ 
take the forms 
$$ L^*= \frac{1}{2}\sum \mathcal{\ N}^2 {\varkappa_D}_{ij} 
q_i^{\prime}q_j^{\prime}-V(q) \quad \text{and} \quad 
H^*=\frac12\sum \frac{1}{\mathcal{\ N}^2} \varkappa_D^{ij} p_i 
p_j+V(q), 
$$ 
respectively. We look for a factor $\N(q)$ such that after the 
above time substitution the equations (\ref{2.1}) take the form 
\begin{equation} 
q_i'=\frac{\partial H^*}{\partial p_i}, \qquad 
p_i'= -\frac{\partial H^*}{\partial q_i}, \qquad i=1,\dots,k. 
\label{x2} 
\end{equation} 
That is, they become Hamiltonian with respect to the symplectic 
form $\varOmega=\sum dp_i\wedge dq_i=\N(q)\Omega+\sum_i P_i 
d\N\wedge dq_i$, where $\Omega=\sum dP_i\wedge dq_i$ is the 
canonical symplectic form on $T^*Q$. In nonholonomic mechanics the 
factor $\N$ is known as the {\it reducing multiplier}. 
 
It appears that non-existence of an invariant measure 
of the reduced system (\ref{2.1}) is an obstruction to its reducibility to a Hamiltonian form. 
Namely, suppose that the original system (\ref{2.1}) is transformed to the Hamiltonian form with 
a reducing multiplier $\N$. Then the system has the 
invariant measure $\N(q)^{k-1}\Omega^k$ (see \cite{St, FeJo, CCLM}). 
According to the celebrated {\it Chaplygin's reducibility 
theorem\/} (see \cite{Ch, Ch2, FeJo} or section III.12 in \cite{NeFu}), for 
$k=2$, the above statement can also be inverted: the existence of 
the invariant measure with the density $\N(q)$ implies that in the 
new time $d\tau=\N(q)dt$, the system (\ref{2.1}) gets the 
Hamiltonian form (\ref{x2}). 
 
Necessary and sufficient conditions for the existence of an 
invariant measure of the reduced system are given by Cantrijn, 
Cortes, de Leon and  Martin de Diego \cite{CCLM}. Recently, a 
nontrivial example of a nonholonomic system (multidimensional 
generalization of the Veselova problem on noholonomic rigid body 
motion \cite{VeVe1, FeKo}) for which the Chaplygin reducibility 
theorem is applicable for any dimension is given by Fedorov and 
Jovanovi\' c \cite{FeJo}. 
 
Note that there is an alternative (but equivalent) description 
of the method of reducing multiplier. The system (\ref{2.1}) can be written as 
$\Omega_{nh}(X_H,\cdot)=dH(\cdot)$,  where $\Omega_{nh}=\Omega+\Xi$ 
is a nondegenerate 2-form. Then the 
Chaplygin multiplier is a function $\N$ such that the form 
$\N\Omega_{nh}$ is closed (see \cite{St, CCLM, EKR}). 
 
\subsection*{Contents of the Paper}
 
In Section 2 we introduce the notion of 
quasi-Chaplygin systems. 
In Section 3 we give a brief description of the Suslov problem and 
show that it can be considered as a quasi-Chaplygin system. 
Furthermore, this provides an example which can be reduced to a 
Hamiltonian form via Chaplygin reducing multiplier. In this sense, 
the complete integrability of the reduced Suslov problem can be 
defined in the natural way (Section 4). The Hamiltonian 
description explains the solvability of multidimensional 
Kharlamova, Klebsh--Tisserand and Lagrange cases obtained in 
\cite{Jo4}. The topology of invariant manifolds of the Kharlamova 
and Klebsh--Tisserand cases is studied in Section 5. Finally, we 
note that the Lagrange case can be treated as a generalized 
Chaplygin system in two different ways. In the first approach the 
reduced system becomes Hamiltonian after the time rescaling, while 
in the second approach the reduced system is already an integrable 
Hamiltonian system, namely the multidimensional spherical 
pendulum.

\section{Quasi-Chaplygin Systems} 

Several  nonholonomic mechanical systems 
have classically been regarded as Chaplygin systems in 
certain properly chosen local coordinates 
(an example is the well known Chaplygin skate), 
although, globally, they are not Chaplygin systems in the 
sense of the above definition (e.g., see \cite{NeFu}). 
 
\begin{definition}{\rm 
With the above notation, 
we say that $(N,l,D,\mathfrak G)$ 
is a {\it quasi-Chaplygin} system, if we allow that 
the sum $D_x+V_x$ does not span the tangent space $T_x N$ on 
some $\mathfrak G$-invariant subvariety $S\subset N$. 
}\end{definition} 
 
As for the Chaplygin systems, the Lagrange--d'Alembert equation 
(\ref{Lagrange}) is $\mathfrak G$-invariant and reduces to the 
quotient space $D/\mathfrak G$ (e.g., see \cite{Ma}). The later 
has a structure of the $\R^k$-vector bundle 
\begin{equation*} 
\begin{array}{ccc} 
\R^k & \longrightarrow & D/\mathfrak G   \\ 
&  & \downarrow  \\ 
&  & Q=N/\mathfrak G 
\end{array}, 
\end{equation*} 
which is not, in general, diffeomorphic to $TQ$. Outside 
$N\setminus S$, we can treat the system as a usual Chaplygin 
system that reduces to $ (D\vert_{N\setminus S}) /\mathfrak G 
\approx T(Q\setminus (S/\mathfrak G)). $ The jumping in the rank 
of the distribution $D_x+V_x$ leads to several interesting 
properties of the system (see examples given below). 
 
Locally, the reduced system can be derived by the use of 
Poincar\'e--Chetayev (or Bolzano--Hamel) equations. Consider an 
open $\mathfrak G$-invariant set $U\subset N$ with local 
coordinates $x=(q,g)$, in which the $\mathfrak G$-action is simply 
$a\cdot (q,g)=(q,ag)$, $a\in \mathfrak G$. The Lagrangian $l$ is 
$\mathfrak G$-invariant. Whence $l(x,\dot x)=l(q,\dot q, 
g^{-1}\dot g)$. 
 
Let $X_1,\dots,X_n$ be linearly independent $\mathfrak 
G$-invariant vector fields on $U$. Then the commutators 
$[X_i,X_j]$ can be written in the basis $X_1,\dots,X_n$: 
$[X_i,X_j]=\sum_k c^k_{ij} X_k$, where structural coefficients 
$c^k_{ij}=c^k_{ij}(q)$ are $\mathfrak G$-invariant functions. Let 
$\omega_1,\dots,\omega_n$ be the {\it quasi-velocities} defined by 
$\dot x=(\dot q,\dot g)=\sum_i \omega_i X_i$. From the definition 
of $\omega_i$, one get the relations $\dot q_i=\sum_{j=1}^n 
A_{ij}\omega_j$, where coefficients $A_{ij}=A_{ij}(q)$ are also 
$\mathfrak G$-invariant functions. 
 
Now, write the Lagrangian as a function of $q$ and $\omega$: $\hat 
l(q,\omega)=l(q,\dot q,g^{-1}\dot g)$. Further, suppose that the 
distribution $D$ is spanned by $X_1,\dots,X_k$. Then, the 
constraints are 
$$\omega_{k+1}=0,\dots,\omega_n=0
$$ 
and 
$\{q_1,\dots,q_k,\omega_1,\dots,\omega_k\}$ can be regarded as 
local coordinates on $D/\mathfrak G$. With the above notation, the 
Poincar\'e--Chetayev equations of the system (e.g., see 
\cite{NeFu, EKR}): 
$$ 
\frac{d}{dt}\left(\frac{\partial \hat l}{\partial 
\omega_i}\right)= \sum_{l=1}^n\sum_{j=1}^k c^l_{ji}(q) 
\frac{\partial \hat l}{\partial \omega_l}\omega_j+X_i(\hat l), 
\quad i=1,\dots,k, 
$$ 
together with the kinematic equations $\dot q_i = \sum_{j=1}^k 
A_{ij}(q) \omega_j$, form a closed system in variables 
$\{q_1,\dots,q_k,\omega_1,\dots,\omega_k\}$. Note that, contrary 
to Chaplygin systems, the matrix $(A_{ij}(q))_{1\le i,j \le k}$ 
does not need to be invertible for all $q$.

\section{Suslov Problem as a Quasi-Chaplygin System} 
 
\subsection*{Suslov Problem} 

Consider the motion of an $n$-dimensional 
rigid body around a fixed point $O$ in the $n$-dimensional 
Euclidean vector space $(\mathcal V,(\cdot,\cdot))$. Let 
$\mathfrak E_1,\dots,\mathfrak E_n$ be the orthonormal frame fixed 
in the body and $\mathfrak e_1,\dots,\mathfrak e_n$ be the 
orthonormal frame fixed in the space. The configuration space of 
the system is the Lie group $SO(n)$: the element $g\in SO(n)$ maps 
the moving coordinate system to the fixed one. We use the 
following usual matrix notation. Let 
$$ 
E_1=(1,0,\dots,0)^t,\; \dots,\; E_n=(0,\dots,0,1)^t. 
$$ 
We take $\{\mathfrak E_1,\dots,\mathfrak E_n\}$ for the base of 
$\mathcal V$. Then $E_1,\dots,E_n$ and 
$$ 
e_1=(e_{11},\dots,e_{1n})^t, \dots,e_n=(e_{n1},\dots,e_{nn})^t,\quad 
e_{ij}=(\mathfrak e_i,\mathfrak E_j) 
$$ 
will be 
the coordinate expressions of $\mathfrak E_1,\dots,\mathfrak E_n$ and of 
$\mathfrak e_1,\dots,\mathfrak e_n$ respectively. The 
matrix $g\in SO(n)$ maps the vectors in the moving frame 
to the same vectors regarded in the fixed frame. Therefore $E_i=g 
\cdot e_i$ and  $g=(e_1,\dots,e_n)^t$, i.e., $g_{ij}=e_{ij}$. 
Note that we can consider the components of the vectors $e_1,\dots, e_n$ 
as redundant coordinates on $SO(n)$. 
 
For a path $g(t)\in SO(n)$, {\it the angular velocity} of the body 
is defined by $\omega(t)=g^{-1}\cdot g(t) \in so(n)$. From the 
conditions $0=\dot E_i=\dot g \cdot e_i+g\cdot \dot e_i$, we find 
that $e_1,\dots,e_n$ satisfy the Poisson equations 
\begin{equation} 
\label{Poisson} 
\dot e_i= -\omega \cdot e_i,\qquad i=1,\dots,n. 
\end{equation} 
 
The kinetic energy of the rigid body is a left-invariant function on $TSO(n)$ 
of the form $\frac12\langle {\mathcal I}\omega,\omega\rangle$, 
where ${\mathcal I}\,:\, so(n)\to so(n)$ in non-degenerate {\it inertia operator} 
and $\langle\cdot,\cdot\rangle$ denotes the Killing metric on $so(n)$. 
For a ``physical'' rigid body, $\mathcal I\omega$ has the form 
$I\omega+\omega I$, where $I$ is a symmetric $n\times n$ matrix called {\it mass tensor} 
(see \cite{FeKo}). 
Further, suppose that the body is placed in a potential field 
that is invariant with respect to the orthogonal transformations which fix $\mathfrak e_n$. 
In our notation this means that the Lagrangian has the form 
\begin{equation} 
l(g,\dot g)=\frac12\langle \mathcal I \omega,\omega\rangle-v(e_n). 
\label{Lagrangian}\end{equation} 
 
The Suslov problem describes the motion of a rigid body with the 
left-invariant constraints (see Fedorov and Kozlov \cite{FeKo}) 
\begin{equation} 
\langle \omega, E_i\wedge E_j\rangle=0, \quad 1 \le i<j\le n-1.  \label{constraints} 
\end{equation} 
 
Equivalently, we can say that the velocity $\dot g$ belong to the left-invariant 
distribution $D_g=g\cdot\D \subset T_g SO(n)$, where 
$$ 
\D=\Span\{E_1\wedge E_n,\dots,E_{n-1}\wedge E_n\}. 
$$ 
 
The motion of the system is described by the Lagrange--d'Alembert 
equation (\ref{Lagrange}), which in the left trivialization takes the form of the 
Euler--Poincar\'e--Suslov (EPS) equation (see \cite{FeKo}) 
\begin{equation} 
\left< \frac{d}{dt} \left({\mathcal I}\omega\right)-[\mathcal I\omega,\omega ]- 
\frac{\partial v}{\partial e_n}\wedge e_n,\eta\right>=0, 
\quad \mathrm{for\;all} \quad \eta\in \D. 
\label{Suslov_eq} 
\end{equation} 
 
The EPS equation, together with the Poisson equations (\ref{Poisson}) and the constraints 
(\ref{constraints}) completely describe the motion 
of the Suslov problem in the variables $\{e_1,\dots,e_n,\omega\}$. 
For $n=3$ these are the equations of the classical Suslov problem 
with the following nonholonomic constrain: the projection of the angular velocity 
to the vector $e_3$ equals zero (see \cite{Su, Koz1, AKN}). 
 
\subsection*{Geometry of the Constraints} 

Let $H\cong SO(n-1)$ be the subgroup of $SO(n)$ with the Lie algebra 
$\mathfrak h=\D^\perp\cong so(n-1)$: 
$$ 
\left( \begin{array}{cc}   SO(n-1) & \mathbf{0}^t \\ 
                     \mathbf{0} & 1 \\ 
               \end{array} \right), \quad \mathbf{0}=(0,\dots,0). 
$$ 
 
Through the paper we shall simply write $SO(n-1)$ instead of $H$. 
We have the following simple geometrical lemma. 
 
\begin{lem}\label{geometric} 
The left action of $SO(n-1)$ 
represents the rotations of a body around the vector $\mathfrak e_n$ fixed in the space, 
while the right action represents the rotations around 
the vector $\mathfrak E_n$ fixed in the body. 
\end{lem} 
 
Consider the {\it left} action of $SO(n-1)$ on $SO(n)$ and the principal bundle 
\begin{equation} 
\begin{array}{cccc} 
SO(n-1) & \longrightarrow & SO(n) &  \\ 
&  & \downarrow & \pi \\ 
&  & S^{n-1}= SO(n)/SO(n-1) & 
\end{array}. 
\label{bundle} 
\end{equation} 
According to Lemma \ref{geometric},  the sphere $S^{n-1}$ (the 
space of cosets $SO(n-1)\cdot g$) can be identified with the 
vector $e_n$ via bijection $e_n \longleftrightarrow SO(n-1)\cdot(e_1,\dots,e_n)^t. $ 
 
The distribution $D\subset TSO(n)$ is invariant 
with respect to the left $SO(n-1)$-action on $TSO(n)$, but 
it cannot be regarded as a connection of 
the principal bundle (\ref{bundle}). 
However $D$ can be seen as a connection almost everywhere on $SO(n)$. 
 
\begin{thm} \label{almost_Chaplygin} 
The distribution $D$ (\ref{constraints}) 
can be regarded as a principal connection of the 
bundle (\ref{bundle}) outside the submanifold $\{e_{nn}=0\}$. 
\end{thm} 
 
\noindent{\it Proof.} 
Since we deal with the left $SO(n-1)$-action, 
the vertical distribution $V$ is right invariant: $V_g=so(n-1)\cdot g$. 
We shall prove that $D_g$ and $V_g$ span the tangent space at $g$ 
outside the submanifold $\{e_{nn}=0\}$. 
 
We have: 
\begin{eqnarray} 
D_g+V_g \ne T_g SO(n) 
&\Longleftrightarrow & \D+g^{-1}\cdot so(n-1)\cdot g \ne so(n) \nonumber\\ 
&\Longleftrightarrow & g\cdot \D \cdot g^{-1} + so(n-1) \ne so(n)\nonumber\\ 
&\Longleftrightarrow & g\cdot X\cdot g^{-1}\in so(n-1), \quad \text{for some} \quad X\in\D. 
\label{not_span} 
\end{eqnarray} 
 
Suppose that there is a vector $X\in\D$ such that $g\cdot X\cdot g^{-1}\in so(n-1)$. 
Let $\bar g=g^{-1}$, $E_i=\bar g \cdot \bar e_i$. 
If $X=x_1 E_1\wedge E_n+\dots+x_{n-1} E_{n-1}\wedge E_n$, then, by using 
$\bar g^{-1}\cdot E_i\wedge E_j \cdot \bar g=\bar e_i \wedge \bar e_j$, we get 
$$ 
\bar g^{-1} \cdot X \cdot \bar g=x_1 \bar e_1\wedge \bar e_n+\dots+ 
x_{n-1} \bar e_{n-1}\wedge \bar e_n= Y\wedge \bar e_n \in so(n-1), 
$$ 
where $Y=x_1 \bar e_1+\dots+x_{n-1} \bar e_{n-1}$. 
On the other hand, $Y\wedge \bar e_n$ belongs to $so(n-1)$ if and only if 
\begin{equation} 
Y_n=x_1 \bar e_{1n}+\dots+x_{n-1} \bar e_{n-1,n}=0 \quad \mbox{and} 
\quad \bar e_{nn}=0. 
\label{Y} 
\end{equation} 
But $\bar e_{nn}=e_{nn}$, hence we proved that (\ref{not_span}) implies $e_{nn}=0$. 
 
To prove the opposite statement, we just note that 
the relation (\ref{Y}) considered as an equation in the variables $x_i$ 
always has a solution. Let us choose $x_i$ such that (\ref{Y}) holds. 
Then, if $\bar e_{nn}=e_{nn}=0$, we conclude that 
the element $X=x_1 E_1\wedge E_n+\dots+x_{n-1} E_{n-1}\wedge E_n\in \D$ 
satisfies $\bar g^{-1} \cdot X \cdot \bar g= g\cdot X \cdot g^{-1}\in so(n-1)$. 
The theorem is proved.  $\Box$ 
 
\subsection*{Reduced Suslov Equation}
 
The Suslov equations (\ref{Suslov_eq}), (\ref{Poisson}), (\ref{constraints}) 
are $SO(n-1)$-invariant and equations 
(\ref{Suslov_eq}), (\ref{constraints}) together with $\dot e_n=-\omega\cdot e_n$ 
can be viewed as their $SO(n-1)$-reduction to 
$$ 
S^{n-1}\times\R^{n-1}\{e_n,\omega\}\approx D/SO(n-1). 
$$ 
 
For the simplicity, denote the vector $e_n$ by $\q=(q_1,\dots,q_n)$. 
Further, suppose that $\I$ preserves the decomposition $so(n)=so(n-1)+\D$. 
Note that $(so(n),so(n-1))$ is a symmetric pair, i.e., $[\D,\D]\subset so(n-1)$. 
Therefore, if $\D$ is an eigenspace of $\I$, then, 
in view of the condition $\pr_\D [\D,\D]=0$, the EPS equation (\ref{Suslov_eq}) 
and the Poisson equation for $e_n$ become 
\begin{eqnarray} 
&& (\I\dot\omega)_{in}=\frac{\partial v}{\partial q_i}q_n-\frac{\partial v}{\partial q_n}q_i, \label{Suslov_eq3}\\ 
&&\dot q_i=-\omega_{in} q_n, \quad i=1,\dots,n-1, \quad 
\dot q_n=\sum_{i=1}^{n-1}\omega_{in} q_i. 
\label{velocities} 
\end{eqnarray} 
 
Contrary to the general case, the {\it reduced Suslov problem} 
(\ref{Suslov_eq3}), (\ref{velocities}) preserves the standard 
measure in the variables 
$\{q_1,\dots,q_n,\omega_{1n},\omega_{2n},\dots,\omega_{n-1,n}\}$. 
By the Euler--Jacobi theorem (e.g., see \cite{AKN}), for the 
integrability (more precisely, solvability by quadratures) of the 
reduced problem (\ref{Suslov_eq3}), (\ref{velocities}), we need 
$2n-5$ additional integrals which are independent of the energy 
integral $\frac12\langle \I\omega,\omega\rangle+v(\q)$. In 
particular, for $n=3$ we need only one additional integral (see 
\cite{KZ, Koz1, AKN}). 
 
However, as it is shown in \cite{Jo4}, we do not need $2n-5$ 
integrals to solve the multidimensional variants of Kharlamova, 
Klebsh--Tisserand and Lagrange cases of the Suslov problem. Below 
we will give the natural definition of a complete integrability of 
(\ref{Suslov_eq3}), (\ref{velocities}) such that the examples 
studied in \cite{Jo4} provide the simplest completely integrable 
cases. Note that, in general, it is still not clear how to define 
the notion of a complete integrability (in the sense of the 
Liouville theorem) of nonholonomic systems (see \cite{Koz1, 
BaCu}). 
 
\subsection*{Chaplygin Reduction}
 
The Lagrangian function $l$ and the distribution $D$ 
are both invariant with respect to the left $SO(n-1)$-action on $TSO(n)$. 
As follows from Theorem \ref{almost_Chaplygin}, the multidimensional 
Suslov problem is a quasi-Chaplygin system. 
Furtermore, after the appropriate time rescaling, the system becomes Hamiltonian. 
 
The reduced space $(SO(n)\setminus\{e_{nn}=0\})/SO(n-1)$ is the union 
of two half-spheres with $q_n>0$ and $q_n<0$. 
On the half-spheres we can use coordinates $q=(q_1,\dots,q_{n-1})$ within the ball 
$$ 
B=\{q\in \R^{n-1}\, \vert\,   (q,q)=q_1^2+\dots+q_{n-1}^2<1\}. 
$$ 
 
Using (\ref{Lagrangian}) and (\ref{velocities}) we can write 
down the reduced Lagrangian and Hamiltonian 
\begin{equation} 
L_\pm=\frac{1}{2}\frac{1}{1-(q,q)}(J\dot q,\dot q) -V_\pm(q), 
 \quad H_\pm=\frac12(1-(q,q))(AP,P)+V_\pm(q), 
\label{L} 
\end{equation} 
where $V_\pm(q)=v(q,q_n)$, $q_n=\pm\sqrt{1-(q,q)}$, the metric 
$J$ is given by $J_{ij}=\I_{in,jn}$, $P=\frac{1}{q_n^2} J\dot q$ and $A=J^{-1}$. 
 
We do not need to find the curvature of the 
connection to find the reduced system. In view of 
(\ref{Suslov_eq3}), after straightforward computations, we can 
get: 
 
\begin{thm} 
\label{multiplier} Suppose that the inertia operator $\I$ 
preserves the decomposition $so(n)=so(n-1)+\D$. After the 
Chaplygin reduction outside submanifold $\{e_{nn}=0\}$, the 
multidimensional Suslov problem (\ref{Suslov_eq}), 
(\ref{Poisson}), (\ref{constraints}) takes the following form on 
$T^*B$ 
\begin{equation} 
\dot q=\frac{\partial H_\pm}{\partial P},\quad \dot 
P=-\frac{\partial H_\pm}{\partial q}+(AP,q)P-(P,AP)q. \label{EL1} 
\end{equation} 
 
Under the time substitution $d\tau =q_n dt$ and an appropriate change of 
momenta $p=q_n P$, the reduced system (\ref{EL1}) 
becomes a Hamiltonian system describing a motion of a particle 
within the ball $B$, 
\begin{equation} 
q'=\frac{\partial H^*_\pm}{\partial p}=Ap, \quad p'= 
-\frac{\partial H^*_\pm}{\partial q}=-\frac{\partial V}{\partial 
q} \quad \Longleftrightarrow  \quad Jq''=-\frac{\partial 
V}{\partial q}, \label{EL2} 
\end{equation} 
where $H^*_\pm(q,p)=\frac12(Ap,p)+V_\pm(q)$. 
\end{thm} 
 
\subsection*{LL Systems as a Quasi-Chaplygin Systems}
 
We can consider Suslov-type problems on other Lie groups as 
nonholonomic systems with left-invariant Lagrangians and left-invariant nonintegrable 
distributions (see \cite{Koi, FeJo2} and references therein). 
 Suppose we are given a natural nonholonomic system with a 
left-invariant Lagrangian $l$ and a 
left-invariant distribution $D$ on a Lie group $G$ (so called LL system). 
Let $\D$ be the restriction of $D$ to $\g=T_{Id} G$. 
Futher, suppose that there is a subgroup $H$ of $G$ such that for its Lie algebra 
$\h$ and $\D$ hold $\g=\h + \D$, $\h \cap \D=0$. 
Then LL system is a quasi-Chaplygin system $(G,l,D,H)$ with respect 
to the left action of $H$. The system reduces to the 
$\dim\D$-dimensional vector bundle over the homogeneous space $G/H$. 
As for the Suslov problem, 
the subvariety $S\subset G$ where 
the vertical spaces $V_g=\h\cdot g$ and the distribution 
$D_g=g\cdot\D$ do not span the tangent spaces $T_g G$ 
is given by the equation $g\cdot \D \cdot g^{-1} + \h \ne \g$.

\section{Integrability} 

 In view of (\ref{velocities}), (\ref{EL2}), we can identify 
\begin{equation} 
\label{identification} 
-\omega_{in} \longleftrightarrow (Ap)_i, \quad 
i=1,\dots,n-1. 
\end{equation} 
and  write (\ref{Suslov_eq3}), (\ref{velocities}) in the form 
\begin{eqnarray} 
&&\dot p=-q_n\frac{\partial v}{\partial q}+ 
\frac{\partial v}{\partial q_n}q, \nonumber\\ 
&&\dot q=q_n Ap,\quad \dot q_n=-(Ap,q).\label{Suslov_eq4} 
\end{eqnarray} 
 
The phase space $\mathcal M=S^{n-1}\{\q\}\times\R^{n-1}\{p\}$ of 
the reduced Suslov problem (\ref{Suslov_eq4}) has the following natural decomposition 
\begin{eqnarray*} 
&\mathcal M=S^{n-1}\{\q\}\times\R^{n-1}\{p\} \approx T^*\bar 
B \cup_\Lambda T^*\bar B, \\ 
&\q=(q,\pm\sqrt{1-(q,q)})\in S^{n-1}  \longleftrightarrow q\in \bar B, 
\end{eqnarray*} 
where $\Lambda$ be the boundary of $T^*B$. 
 
According to Theorem \ref{multiplier}, in the new time $\tau$ and 
outside the domain $\Lambda$ the dynamics is Hamiltonian. 
Note that for $q_n>0$, time $\tau$ has the same direction as 
$t$, while  for $q_n<0$ the direction is opposite. 
Therefore, the closure of flows of Hamiltonian vector fields 
$X_{H^*_+}$ and $X_{H^*_-}$ to $T^*\bar B$ 
recovers the dynamic of (\ref{Suslov_eq4}) 
in the following manner. Let 
$$ 
\Sigma=\left\{(q,p)\in\Lambda\,\vert\, (Ap,q)=0, \, 
\frac{\partial v}{\partial q_n}\bigg\vert_{\q=(q,0)}=0\right\} 
$$ 
be the set of equilibria points of the system (\ref{Suslov_eq4}) with $q_n=0$. 
 
Let $(q_0,p_0)\in\Lambda\setminus\Sigma$  and let 
$(q_+(\tau),p_+(\tau))$  and $(q_-(\tau),p_-(\tau))$ 
be the trajectories of 
Hamiltonian vector fields $X_{H^*_+}$ and $X_{H^*_-}$ such that 
$$ 
\lim_{\tau\to+\tau_0}(q_+(\tau),p_+(\tau))= 
\lim_{\tau\to-\tau_0}(q_-(\tau),p_-(\tau))=(q_0,p_0). 
$$ 
Then, invrting the quadrature 
$$ 
\int_{\tau_{0}}^{\tau}\frac{\pm d\tau}{\sqrt{1-(q_\pm(\tau),q_\pm(\tau))}}=t-t_0, 
$$ 
we find $t=t(\tau)$ and the dynamics of $\q=(q,q_n)$ in the original time $t$. 
 
The points $(q,p)\in\Sigma$ that correspond to equilibrium points of 
(\ref{Suslov_eq4}) are reached in an infinite time $t$. They not need to be 
the equilibrium points of vector fields $X_{H^*_+}$ and $X_{H^*_-}$. 
 
\begin{definition}\label{def} 
{\rm 
We shall say that the reduced Suslov problem (\ref{Suslov_eq4}), 
is {\it completely integrable} if one can find 
$n-1$ independent smooth integrals $f_i: \mathcal M\to\R$, which after 
substitutions $q_n=\pm\sqrt{1-(q,q)}$ Poisson commute between 
themselves.} 
\end{definition} 
 
\epsfbox{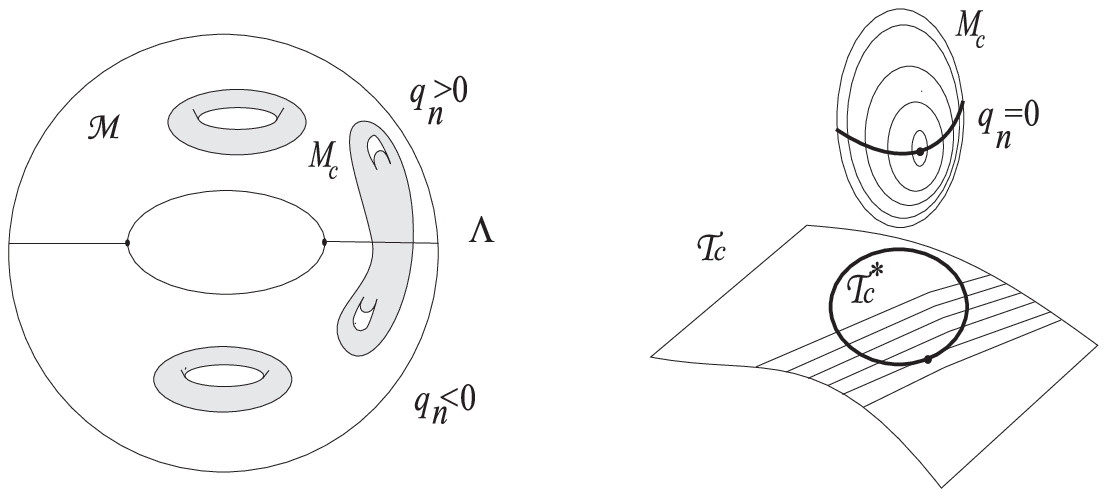} 
 
\centerline{Figure 1. (a) Complete integrability $\qquad$ (b) 
Kharlamova case} 
 
\medskip
 
In particular, we can take $f_1=\frac12(Ap,p)+v(\q)$. 
In the case of integrability, due to the Liouville 
theorem, one can (locally) find trajectories 
$(q_\pm(\tau),p_\pm(\tau))$ by quadratures (e.g., see \cite{AKN}). 
On the other hand, the topological part of the theorem can not be 
directly applied. Namely, if a compact connected component $\mathcal L_c$ of the 
regular invariant set 
\begin{equation} 
\mathcal M_c=\{(\q,p)\in \mathcal M \,\vert\, f_i(\q,p)=c_i, \; i=1,\dots,n-1\} 
\label{invariant_set} 
\end{equation} 
does not intersect $\Lambda$, then $\mathcal L_c$ is an 
$(n-1)$-dimensional torus with a uniform quasi-periodic dynamics 
in $\tau$ and therefore with a non-uniform quasi-periodic motion 
in the original time $t$. However, if $\mathcal L_c$ intersect 
$\Lambda$, it may have a quite complicate topology (as an 
illustration, see Figure 1a). In the three-dimensional case, the 
topological structure of invariant manifolds of several integrable 
variants of the Suslov problem was studied by Tatarinov \cite{Ta} 
and Okuneva \cite{Ok, Ok2}.

\subsection*{Reconstruction}
 
To reconstruct the motion $(g(t),\dot g(t))$ on the whole phase 
space $D$, we have to solve the Poisson equations (\ref{Poisson}) 
for $e_1,\dots,e_{n-1}$, i.e., to find all trajectories in $D$ 
that projects to the given trajectory 
$(e_n(t),\omega(t))=(\q(t),Ap(t))$. If $(e_n(t_0),\omega(t_0))$ is 
an equilibrium point or if $(e_n(t),\omega(t))$ is a periodic 
orbit, then the invariant set 
$\pi^{-1}\{(e_n(t),\omega(t))\}\subset D$ is foliated with 
invariant tori of maximal dimension $\mathrm{rank}\,SO(n-1)$ or 
$\mathrm{rank}\,SO(n-1)+1$, respectively (e.g., see Zenkov and 
Bloch \cite{Bl_Z} and references therein). 
 
\section{Examples of Topology of Invariant Manifolds} 
 
\subsection*{Systems with the Time Symmetry} 
 
Suppose that the potential satisfy the condition 
$V_+(q)=V_-(q)=V(q)$. Then the reduced Suslov equations 
(\ref{Suslov_eq3}), (\ref{velocities}), i.e., (\ref{Suslov_eq4}), 
are invariant with respect to the transformation $t\mapsto -t$, 
$q_n\to-q_n$, and $\Sigma$ is simply given by 
$$ 
\Sigma=\{(p,q)\in\Lambda \; \vert \; (Ap,q)=0\}. 
$$ 
 
Thus, in the time $t$, 
we are going along the trajectory $(q_+(\tau),p_+(\tau))$ of 
$X_{H^*_+}$ until we reach $\Lambda\setminus \Sigma$. Then we 
continue to go along the same trajectory, but in the opposite 
direction. If we reach again the boundary $\Lambda\setminus 
\Sigma$, we get a closed trajectory. 
 
As an example, consider the Suslov problem with a rigid 
body inertial operator $\I\omega=I\omega+\omega I$, 
$I=\diag(I_1,\dots,I_n)$ and a quadratic potential $v(\q)=C_1 
q_1+\dots+C_{n-1} q_{n-1}+ \frac12(B_1 q_{1}^2+\dots+B_n 
q_{n}^2)$. Then 
\begin{eqnarray} 
&&H^*= H^*_\pm=\frac12\left(\frac{1}{I_1+I_n}p_1^2+\dots+ 
\frac{1}{I_{n-1}+I_{n}}p_{n-1}^2\right)+\nonumber\\
&&+C_1 q_1+\dots+C_{n-1} q_{n-1}+\frac12\left((B_1-B_n) q_{1}^2+\dots+(B_{n-1}-B_n) q_{n-1}^2\right). 
\label{H^*} 
\end{eqnarray} 
 
The functions 
$$ 
f_i(q,p)=p_i^2+2C_i (I_i+I_n)q_i+(I_i+I_n)(B_i-B_n)q_i^2, \qquad i=1,\dots,n-1 
$$ 
Poisson commute and they also commute with the Hamiltonian (\ref{H^*}). 
Moreover, they are smooth functions on $\mathcal M$. Thus, the system 
(\ref{Suslov_eq4}) is completely integrable. 
 
Furthermore, we can consider the flow of $X_{H^*}$ and integrals $f_i$ on the 
whole symplectic linear space $\R^{n-1}\{q\}\times\R^{n-1}\{p\}$. 
Let $\T_c=\{f_i=c_i\}\subset\R^{2n-2}$ be an invariant 
submanifold of $X_{H^*}$. Then the invariant set 
$\mathcal M_c=\{f_i=c_i\}\subset \mathcal M$ is a two-fold covering of 
$$ 
\T^*_c=\T_c \cap \{q_1^2+\dots+q_{n-1}^2\le 1\} 
$$ 
determined by the multivalued function $q_n=\pm \sqrt{1-(q,q)}$. 
The branching points of the covering correspond to zeros of 
$q_n$, i.e., to $\partial \T^*_c \approx\Lambda \cap \mathcal M_c$. 
 
\subsection*{The Kharlamova Case}
 
Let $v(\q)=C_1 q_1+\dots+C_{n-1} q_{n-1}$ (multidimensional {\it Kharlamova case}). This 
is a potential for a rigid body placed in a homogeneous gravitation 
field with the mass center orthogonal to  $\mathfrak E_n$.  The trajectories of the system can be found by 
quadratures and can be expressed in terms of elliptic functions of 
time $t$ (see Jovanovi\' c \cite{Jo4}). 
 
A generic invariant set $\T_c$ is diffeomorphic to $\R^{n-1}$ and 
$\T^*_c$ is a disjoint union of $2^l$ copies of 
$(n-1)$-dimensional closed balls (the number $l$, $0\le l \le 
n-1$ depends on the choice of constants $c_i$). Therefore, connected 
components of $\mathcal M_c$ are spheres. 
Trajectories of the vector field $X_{H^*}$ pass through $\Lambda$. 
Therefore, apart from a finite number of equilibrium points and their 
asymptotic trajectories, all the trajectories of the Suslov problem in 
the original time $t$ will be closed (Figure 1b). Thus we get 
 
\begin{thm} 
In the Kharlamova case, the phase space $\mathcal M$ of the reduced problem 
 is almost everywhere foliated with $(n-1)$-dimensional 
spheres. The distribution $D$ is filled up with 
conditionally-periodic trajectories of maximal dimension 
$[n/(n-1)]+1$. 
\end{thm} 
 
\subsection*{Klebsh--Tisserand Case}
 
For generic values of the constants $C_i$ and $B_i$ the topology 
of invariant manifolds is much more complicated. Consider the 
quadratic potential $v(\q)=\frac12(B_1 q_{1}^2+\dots+B_n q_{n}^2)$ 
(multidimensional {\it Klebsh--Tisserand case} \cite{Jo4}) with 
$B_i>B_n$, $i=1,\dots,n-1$. Then $\T_c$ is the $(n-1)$-dimensional 
torus $S_{c_1}\times\dots\times S_{c_{n-1}}$, where 
$S_{c_i}=\{f_i=c_i\}$ are circles in the planes $\R^2\{p_i,q_i\}$. 
(If some of $c_i$ vanish, then the dimension of the tori 
decreases.) Let $\varphi_i$ be angular variables on the circles 
$S_{c_i}$: 
$$ 
p_i=\sqrt{c_i}\cos\varphi_i, \quad 
q_i=\sqrt{\frac{c_i}{\varkappa_i}}\sin\varphi_i, \quad 
\varkappa_i={(I_i+I_n)(B_i-B_n)}, \qquad i=1,\dots,n-1, 
$$ 
Then the Hamilton equations on $\T_c$ take the form 
\begin{equation} 
\frac{d\varphi_i}{d\tau}=\Omega_i=\sqrt{\frac{B_{i}-B_n}{I_{i}+I_n}}, 
\qquad i=1,\dots,n-1 \label{torus_eq} 
\end{equation} 
and the subset $\T^*_c\subset \T_c$ is given by equation 
\begin{equation} 
\frac{c_1}{\varkappa_1}\sin^2\varphi_1+\dots 
+\frac{c_{n-1}}{\varkappa_{n-1}}\sin^2\varphi_{n-1}\le 1. 
\label{condition} 
\end{equation} 
 
Using the relation (\ref{condition}), we can describe the 
topological structure of the invariant manifolds $\mathcal M_c$ of the 
Klebsh--Tisserand case. Namely, the condition 
\begin{equation} \label{cond1} 
\frac{c_1}{\varkappa_1}+\dots+\frac{c_{n-1}}{\varkappa_{n-1}}<1 
\end{equation} 
implies that $q_n\ne 0$ on $\T_c$. As a result, the following theorem holds 
 (see \cite{Jo4}). 
 
\begin{thm} \label{Klebsh} 
In the domain defined by the condition (\ref{cond1}), 
invariant manifolds (\ref{invariant_set}) are 
disjoin unions of two $l$-dimensional tori ($l\le n-1$) 
with non-uniform conditionally-periodic motions. 
\end{thm}

For $\frac{c_1}{\varkappa_1}+\dots+\frac{c_{n-1}}{\varkappa_{n-1}}\ge 1$, the boundary 
of $\T_c^*$ is not empty and there are 
several cases depending of the values of $c_i$. We quote some of them.

\begin{itemize} 
\item[(i)] $0< c_1 < \varkappa_1,\dots, 0< c_{n-1} <\varkappa_{n-1}$. 
 
In this case $\T^*_c$ is obtained by removing a disjoint union 
 
of $2^{n-1}$ open balls $\mathcal B_i$ from the torus $\T_c$. The 
invariant set $\mathcal M_c$ is diffeomorphic to 
$$ 
(\T^{n-1}_c\setminus \cup_i \mathcal B_i)\amalg _{\cup_i 
\partial \mathcal B_i} (\T_c^{n-1}\setminus \cup_i \mathcal B_i). 
$$ 
For $n=3$ this manifold is a sphere with five handles (see Figure 2). 
 
\item[(ii)] $c_1>\varkappa_1,\dots,c_{n-1}>\varkappa_{n-1}$. 
 
$\T^*_c$ is a disjoint union of $2^{n-1}$ closed balls. 
Similarly to the Kharlamova case, $\mathcal M_c$ is 
diffeomorphic to disjoint union of $2^{n-1}$ spheres $S^{n-1}$. 
 
\item[(iii)] $c_1>\varkappa_1, 
0< c_2<\varkappa_2,\dots,0< c_{n-1}<\varkappa_{n-1}$. 
 
The subset  $\T^*_c$ is diffeomorphic to two copies of 
$\mathbb{T}^{n-2}\times [0,1]$. 
As a result, 
$\mathcal M_c$ is a two-fold covering 
of $2\mathbb{T}^{n-2}\times [0,1]$ 
with branching point on the boundaries $\mathbb{T}^{n-2}\times 
\{0,1\}$, i.e., 
$\mathcal M_c$ is a disjoint union of two $(n-1)$-dimensional tori $\mathbb{T}^{n-1}$ 
(Figure 2). 
\end{itemize} 
 
\medskip 
 
\epsfbox{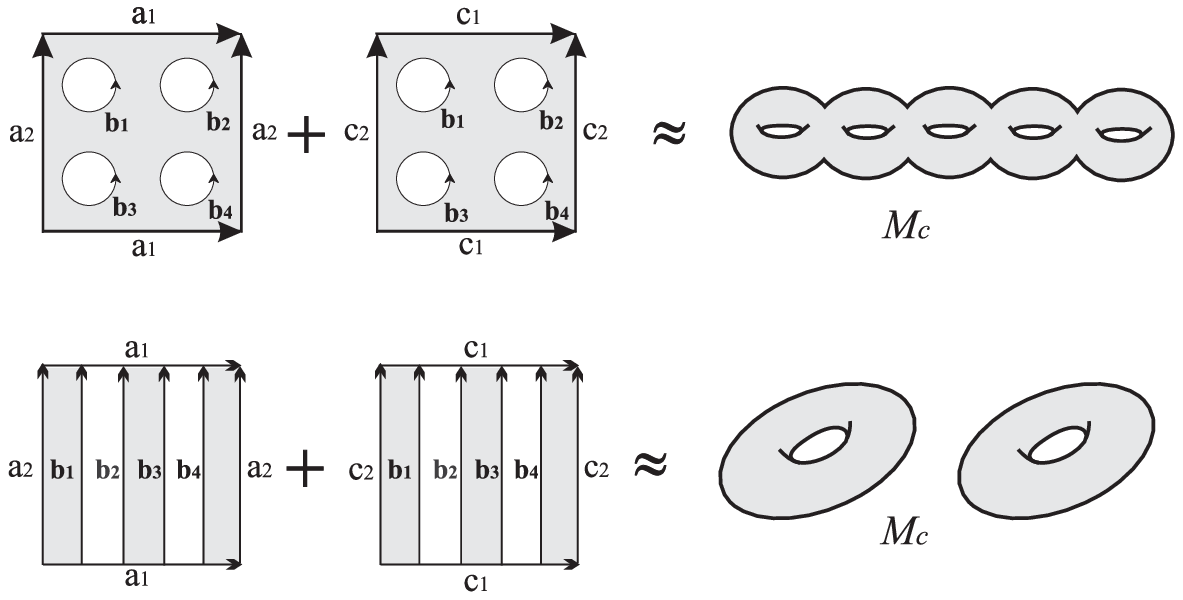} 
 
\centerline{Figure 2. Klebsh--Tisserand Case} 
 
\medskip

Analysis of other cases goes along similar lines. It is desirable 
to have a complete description of the bifurcation diagram of the 
mapping $\mathrm{F}=(f_1,\dots,f_{n-1}): \mathcal M\to \R^{n-1}$, 
as given for $n=3$ in \cite{Ok}. 
 
For generic values of the frequencies $\Omega_i$, the trajectories 
of (\ref{torus_eq}) are dense in $\T_c$ and,  in the domain 
$\frac{c_1}{\varkappa_1}+\dots+\frac{c_{n-1}}{\varkappa_{n-1}} \ge 1$, a 
generic trajectory intersects the boundary $\partial \T^*_c\approx 
\Lambda \cap \mathcal M_c$. Therefore, in the original time $t$ 
almost all the trajectories are closed.  Thus, concerning the 
reconstruction problem, it would be interesting to describe 
qualitative behavior of the motions over invariant sets $\mathcal M_c$, where 
$\frac{c_1}{\varkappa_1}+\dots+\frac{c_{n-1}}{\varkappa_{n-1}}<1$. 
A similar problem for the 
Fedorov--Kozlov integrable case of the Suslov problem (the 
original constraints (\ref{constraints}) are relaxed and 
$v(e_n)\equiv 0$) is studied in \cite{Bl_Z}. 
 
\subsection*{Lagrange Case} 
 
Consider the EPS equations (\ref{Suslov_eq}), (\ref{Poisson}), 
(\ref{constraints}) describing the motion of a dynamically 
symmetric heavy rigid body. Namely, assume that 
$I=\diag(I_1,\dots,I_1,I_n)$ and that the mass center lies on the 
symmetry axis $\mathfrak E_n$.  This gives us a nonholonomic 
version of the multidimensional {\it Lagrange top} considered by 
Beljaev  \cite{Be}.  The Lagrangian of the system has the form 
\begin{equation} 
l(g,\dot g)=I_1\langle \pr_{so(n-1)}\omega,\pr_{so(n-1)}\omega\rangle+ 
\frac12(I_1+I_n)\langle \pr_{\D}\omega ,\pr_{\D}\omega\rangle-\epsilon e_{nn}. 
\label{lagrange_top} 
\end{equation} 
 
The reduced system (\ref{Suslov_eq4}) is integrable in view 
of the Euler--Jacobi theorem and in the sense of Definition 
\ref{def} as well. Namely, the functions $f_{ij}=q_ip_j-q_j p_i: \mathcal M\to\R$ 
are integrals of (\ref{Suslov_eq4}). There are $2n-5$ independent 
integrals of this form, which, together with the energy 
$\frac12(I_1+I_n)(p,p)+\epsilon q_n$, implies that the phase space 
$\mathcal M$ is foliated with two-dimensional invariant 
manifolds. Since the system has an invariant measure, it is 
integrable by the Euler--Jacobi theorem \cite{AKN}. On the other 
hand, the functions $f_{ij}$, $H^*_{\pm}$ ensure 
the non-commutative integrability (and, therefore, 
the usual complete integrability \cite{BJ}) 
of the Hamiltonian flows $X_{H_\pm^*}$ within $T^*B$. 
 
The geometrical meaning of the integrals 
$f_{ij}$ is that the trajectories $q(\tau)$ take place over 
invariant two-dimensional planes, depending on the initial 
conditions. After fixing the plane, the problem becomes the usual 
two-dimensional problem of the motion of a particle in a central 
potential force field. 
 
\subsection*{Reduction to the Spherical Pendulum}
 
The nonconstrained Lagrange top system is completely integrable 
(see \cite{Be}, the Lax pair for the system is given in 
\cite{RS}). It appears that the Suslov problem (\ref{Suslov_eq}), 
(\ref{Poisson}), (\ref{constraints}) can be regarded as a 
subsystem of the Lagrange top. Namely, the Lagrangian (\ref{lagrange_top}) is 
invariant with respect to both left and right actions of the 
Lie subgroup $SO(n-1)$. Let us consider the {\it right} action. 
(Note that $D$ is a principal connection of the bundle  (\ref{bundle}) given 
by the right action of $SO(n-1)$.) The momentum map of the system 
is $\phi(g,\dot g)=2I_1\pr_{so(n-1)}\omega.$ This gives us the conservation law 
$\frac{d}{dt}\pr_{so(n-1)}\omega=0$. In particular, the 
distribution $D=\phi^{-1}(0)$ is an invariant submanifold of the 
Lagrange top system. 
 
Therefore, the right $SO(n-1)$-Chaplygin reduction of the Suslov 
problem to $TS^{n-1}$ coincides with the Lagrange--Routh 
reduction of the Lagrange top for zero value of the momentum mapping 
(see \cite{AKN}, page 87, Theorem  13). 
According to Lemma \ref{geometric}, 
the sphere $S^{n-1}=SO(n)/SO(n-1)$ can be identified with the 
positions of the vector $\mathfrak E_n$ considered in the fixed frame. 
The reduced Lagrangian has the form 
$$ 
L=\frac12(I_1+I_n)(\dot e_{1n}^2+\dots+\dot 
e_{nn}^2)-\epsilon e_{nn}. 
$$ 
Hence, the reduced system is integrable: it is the 
multidimensional spherical pendulum. 
 
\subsection*{Acknowledgments} 
 
The first author (Yu.F.) acknowledges the support of grant BFM 2003-09504-C02-02 of 
Spanish Ministry of Science and Technology. The second author (B.J.) was supported 
by the Serbian Ministry of Science, Project 
"Geometry and Topology of Manifolds and Integrable Dynamical Systems".

\end{document}